\documentclass[letter]{aa}
\usepackage{txfonts}
\usepackage{graphicx}

\def\H2{H$_2$}

\begin{document}
\title{The ALMA view of the protostellar system HH212}
\subtitle{The wind, the cavity, and the disk}
\author{C. Codella \inst{1} \and 
S. Cabrit \inst{2,3} \and
F. Gueth  \inst{4} \and
L. Podio \inst{1} \and
S. Leurini \inst{5} \and
R. Bachiller \inst{6} \and
A. Gusdorf \inst{2} \and
B. Lefloch \inst{3} \and
B. Nisini \inst{7} \and
M. Tafalla \inst{6} \and
W. Yvart \inst{2}}  

\institute{
INAF, Osservatorio Astrofisico di Arcetri, Largo E. Fermi 5,
50125 Firenze, Italy
\and
LERMA, UMR 8112 du CNRS, Observatoire de Paris, \'Ecole  Normale Sup\'erieure, 61 Av. de l'Observatoire, 75014 
Paris, France
\and
UJF-Grenoble1/CNRS-INSU, Institut de
Plan\'etologie et d'Astrophysique de Grenoble (IPAG) UMR 5274,
Grenoble, 38041, France
\and
IRAM, 300 rue de la Piscine, 38406 Saint Martin d'H\`eres, France
\and
Max-Planck-Institut f\"ur Radioastronomie, Auf dem H\"ugel 69, 53121 Bonn, Germany
\and
IGN, Observatorio Astron\'omico Nacional, Alfonso XIII 3, 28014, Madrid, Spain
\and
INAF, Osservatorio Astronomico di Roma, via di Frascati 33, 00040, Monte Porzio Catone, Italy
}

\offprints{C. Codella, \email{codella@arcetri.astro.it}}
\date{Received date; accepted date}

\authorrunning{Codella et al.}
\titlerunning{The protostellar system HH212 as observed by ALMA}

\abstract
{Because it is viewed simply edge-on, the HH212 protostellar system is an ideal laboratory for studying  
the interplay of infall, outflow, and rotation in the earliest stages of low-mass star formation.} 
{We wish to exploit the unmatched combination of high angular resolution, high sensitivity, 
high-imaging fidelity, and spectral coverage 
provided by ALMA to shed light on the complex kinematics of the innermost central regions of HH212.}
{We mapped the inner $10 \arcsec$ (4500 AU) of the HH212 system at $\simeq 0.5\arcsec$ resolution in 
several molecular tracers and in the 850 $\mu$m dust continuum using the ALMA interferometer in band 7 
in the extended configuration of the Early Science Cycle 0 operations.}
{Within a single ALMA spectral set-up, we simultaneously identify all the crucial ingredients known to be
involved in the star formation recipe namely: 
(i) the fast, collimated bipolar SiO jet driven by the protostar, (ii) the large-scale swept-up CO outflow,  
(iii) the flattened rotating and infalling envelope, with bipolar cavities carved by the outflow (in C$^{17}$O(3--2)),
and (iv) a rotating wide-angle flow that fills the cavities and surrounding the axial jet (in C$^{34}$S(7--6)). 
In addition, the compact high-velocity C$^{17}$O emission ($\pm$ 1.9--3.5 km s$^{-1}$ from systemic)  
shows a velocity gradient along the equatorial plane consistent with a rotating disk of $\simeq 0\farcs2$ = 90 AU
around a $\simeq 0.3 \pm 0.1 M_{\rm \odot}$ source. The rotating disk is possibly Keplerian.}
{HH212 is the third Class 0 protostar with possible signatures of a Keplerian disk of radius $\geq 30 AU$.
The warped geometry in our CS data suggests that this large keplerian 
disk might result from misaligned magnetic and rotation axes 
during the collapse phase. 
The wide-angle CS flow suggests that disk winds may be present in this source.} 


\maketitle

\section{Introduction}

Jets from young accreting stars remain one of the most spectacular and enigmatic phenomena in astrophysics. 
Although their exact launch zone is still debated, it is currently accepted that they are powered by
the rotation and accretion energy of the system, and are accelerated or collimated via a magneto-hydrodynamical (MHD) process
 (see e.g. Ferreira et al. 2006; Shang et al. 2007; Pudritz et al. 2007, and references therein). 
MHD-driven jets  
could play a crucial role during the earliest Class 0 phase, 
that is in the star and disk formation process itself. High-resolution MHD simulations
of protostellar collapse with typical amounts of rotation and magnetisation show that MHD outflows
are an unavoidable outcome, and are able to eject 20\% to 50\% of the infalling core gas before it reaches 
the central source (e.g. Ciardi \& Hennebelle 2010). Hence this initial phase of massive MHD ejection may be a 
key agent that can limit the final stellar mass, and explain the low ($\simeq$ 30\%) core-to-star efficiency 
suggested from a comparison of the IMF with the prestellar core mass function (e.g. Andr\'e et al. 2007).
However, the same simulations show that magnetic braking by the outflows and twisted B-fields is so
efficient when the field and the spin axis $\Omega$ are aligned that keplerian disks may be initially suppressed beyond 10 AU (e.g. Price \& Bate 2007, Machida et al. 2011). This is the so-called magnetic-braking catastrophe. 
On the other hand, much larger keplerian disks of 100-150 AU have previously been
reported in two Class 0 sources (L1527 and VLA1623; Tobin et al. 2012; Murillo et al. 2013). 
Such large Keplerian disks might result from 
misaligned $B-\Omega$ configurations or a strong turbulence (Joos et al. 2012, 2013), 
but observational evidence is still lacking.  
A detailed observation and characterisation of a larger sample of Class 0 infall--outflow systems is thus essential
to elucidate the disk formation process, and to clarify the role of jets and outflows in removing angular momentum 
from the system and regulating the final stellar mass.

HH212 is a strikingly bright and symmetric bipolar jet from a Class 0 source in Orion (at 450 pc)
that was first revealed in H$_2$ imaging (Zinnecker et al. 1998). Its innermost regions have 
been extensively studied in CO(2--1), (3--2) and SiO(2--1), (5--4), (8--7) withf the SMA and IRAM PdBI at scales ranging
from $\simeq$ 1$''$-2.5$''$ (Lee et al. 2006, 2007) down 
to $\simeq$ 0.3$''$-0.4$''$ (Codella et al. 2007; Lee et al. 2008; Cabrit et al. 2007, 2012).
These maps revealed a bright bipolar SiO microjet with inner peaks at $\pm$1--2$''$ = 450--900 AU
of the protostar, invisible in H$_2$ due to high extinction (Codella et al. 2007; Lee et al. 2007).
The SiO jet width $\simeq$ 100 AU is remarkably close to atomic jet widths in T Tauri stars,
favouring a universal MHD collimation process (Cabrit et al. 2007).
Interestingly, a flattened rotating envelope in the equator perpendicular to the jet axis was observed
in NH$_3$ with a radius of $\sim$ 7000 AU by Wiseman et al. (2001). 
Subsequent observations in C$^{18}$O and $^{13}$CO  with the SubMillimeter Array, and 
in HCO$^+$ with the Atacama Large Millimeter Array (ALMA) suggest that the rotating envelope 
inside a radius $\sim$ 1000 AU is in free-fall onto a source of mass $\simeq 0.2M_\odot$ 
while conserving angular momentum (Lee et al. 2006, 2014). 
The dust continuum imaged by ALMA is flattened perpendicular to the jet and suggestive of a 
disk of (maximum) radius $\simeq 0.3$ AU (Lee et al. 2014). 
All these findings make HH212 an ideal laboratory for investigating the interplay 
of infall, outflow and rotation in the earliest
stages of the low-mass star-forming process. 

\section{Observations} 

HH212 was observed with ALMA using 24 12-m antennas  
on 2012 December 1 during the Early Science Cycle 0 phase. 
The shortest and longest baselines were about 20 m and 360 m, 
from which we obtained a maximum unfiltred scale of 3$\arcsec$ at 850 $\mu$m.
The C$^{17}$O(3--2), SiO(8--7), and C$^{34}$S(7--6) 
lines\footnote{Spectroscopic parameters were extracted from the Jet
Propulsion Laboratory molecular database (Pickett et al. 1998) and 
the Cologne Database for Molecular Spectroscopy (M\"uller et al. 2005).} at 337061.13 MHz, 347330.63 MHz,
and 337396.69 MHz, respectively, were observed  
using spectral units of 488 kHz (0.43 km s$^{-1}$) resolution.
Calibration was carried out following standard procedures, using quasars J0538--440, J0607--085,
as well as Callisto and Ganymede. 
Spectral line imaging was achieved with the CASA package.
Data analysis was performed using the  GILDAS\footnote{http://www.iram.fr/IRAMFR/GILDAS} package.
Images have a typical clean-beam FWHM of $0\farcs65\times0\farcs47$ (PA = 35$\degr$), and
an rms noise of $\sim$ 1 mJy beam$^{-1}$ for continuum, 
and 3--4 mJy beam$^{-1}$ in the 0.44 km s$^{-1}$ channels.  
Positions are given with respect to the MM1 protostar, 
located at $\alpha({\rm J2000})$ = 05$^h$ 43$^m$ 51$\fs$41, 
$\delta({\rm J2000})$ = --01$\degr$ 02$\arcmin$ 53$\farcs$17, in excellent agreement
with the coordinates derived by Lee et al. (2014) using ALMA. 

\section{Line results and analysis}

Figure~1 compares the emission maps in the 850 $\mu$m continuum, SiO(8--7), C$^{17}$O(3--2), and C$^{34}$S(7--6).  
The combination of these tracers allows us to simultaneously image {\em in a single ALMA spectral set-up} different
ingredients of the star formation process:
(i) a pair of narrow SiO jets launched from the protostar (dust peak), 
(ii) an extended flattened C$^{17}$O envelope around the outflow waist, 
and (iii) biconical C$^{34}$S emission lobes surrounding the jet. 
The first feature is a well-known characteristics of some Class 0 sources
(see e.g. Tafalla et al. (2010) and references therein). 
A full analysis of the new details detected up by ALMA will be presented in a later publication. 
Here, we focus on the latter two features, which were imaged here with unprecedented fidelity and signal-to-noise.  

\begin{figure}
\centerline{\includegraphics[angle=0,width=8.3cm]{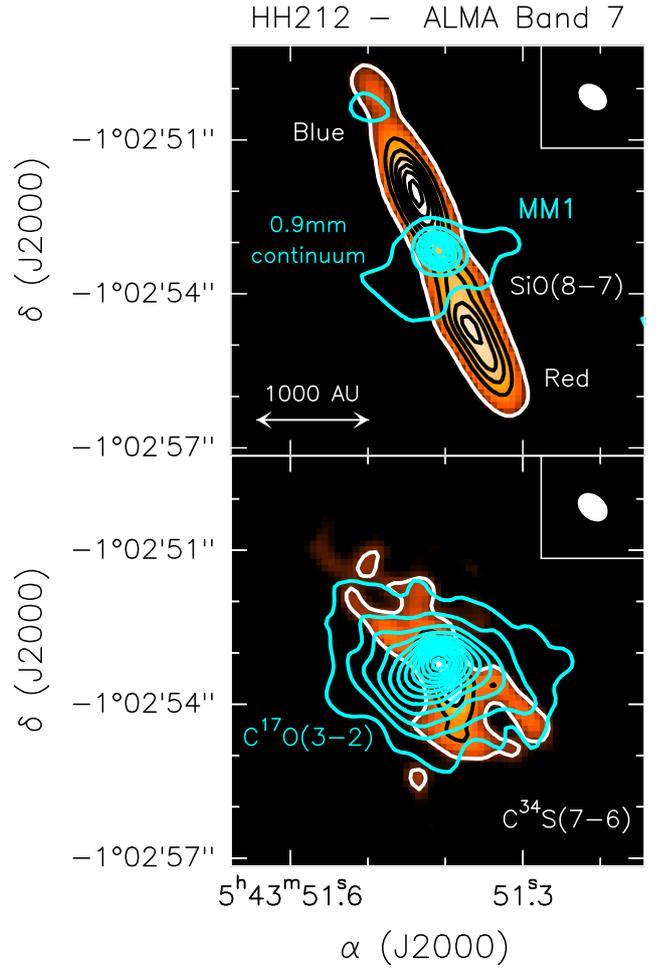}}
\caption{The protostellar system HH212 as observed by ALMA Cycle 0 in band 7.
{\it Upper panel:} 850 $\mu$m continuum (turquoise contours) 
overlaid on SiO(8--7) integrated between --23 and +15 km s$^{-1}$ (colour scale and black contours), 
compared with the systemic velocity $V_{\rm sys}$ (see text). 
First contours at 5$\sigma$ (6 mJy beam $^{-1}$ for
continuum and 270 mJy beam $^{-1}$ km s$^{-1}$ for SiO) in steps of
20$\sigma$ (continuum) and 5$\sigma$ (SiO). 
The ellipse shows the synthesised continuum HPBW ($0\farcs61\times0\farcs45$ at PA = 51$\degr$).
The beam HPBW of the SiO map is $0\farcs63\times0\farcs46$ (PA = 49$\degr$).
{\it Bottom panel:} contour plot of the C$^{17}$O(3--2) emission (turquoise) 
overaid on the of the C$^{34}$S(7--6) emission (colour and black),
both integrated in the $V_{\rm sys}$$\pm$6 km s$^{-1}$ range.
The HPBW is $0\farcs65\times0\farcs47$ (PA = 49$\degr$).
First contours and steps correspond to 5$\sigma$
(50 mJy beam$^{-1}$ km s$^{-1}$ for C$^{34}$S and 65 mJy beam$^{-1}$ km s$^{-1}$ for C$^{17}$O).}
\label{maps}
\end{figure}

\subsection{Systemic velocity}

The channel maps for C$^{17}$O(3--2) and C$^{34}$S(7--6) are reported
in Figs. A.1 and A.2. There is no missing flux feature or P Cygni absorption 
near systemic velocity, in contrast to previous SMA data in the more abundant $^{13}$CO and C$^{18}$O 
molecules and ALMA data in HCO$^{+}$ (Lee et al 2006, 2014). 
The spatial distributions clearly indicate that the velocity where emission is most spatially 
extended and thus presumably closest to the systemic velocity
is in the [+0.91,+1.34] km s$^{-1}$ channel 
for C$^{17}$O, and in the [+1.23,+1.69] km s$^{-1}$ channel for C$^{34}$S
(bottom-right panels of Figs. A.1 and A.2). These are also the channels for which the other channel maps 
show the best blue/red morphological symmetry. 
From combining these findings, and considering our spectral resolution of $\sim$ 0.43 km s$^{-1}$, 
the systemic velocity $V_{\rm sys}$, as indicated by C$^{17}$O and C$^{34}$S, has to be close to $\simeq +1.3\pm0.2$ km s$^{-1}$.
This value is clearly supported by the position-velocity (PV) diagram of the C$^{17}$O(3--2) emission
perpendicular to the jet axis (PA = 112$\degr$), see Fig. 2ab, which
shows that the emission is most extended and most symmetric about
$V_{\rm sys}$ = +1.3 km s$^{-1}$. This value is 
slightly different from the +1.6 km s$^{-1}$ measured using NH$_3$(1,1) VLA emission by Wiseman et al. (2001). 
This could reflect the different spatial scales, as the ammonia core observed by Wiseman et al. (2001) has 
a FWHM $\sim$ 14000 AU and probably traces the motion in the outer layers of the molecular envelope. 
We adopt $V_{\rm sys} = +1.3$ km s$^{-1}$ in the remainder of this paper. 

\subsection{C$^{17}$O(3--2): rotating envelope cavities and inner disk}

Figure 2ab shows that, in addition to infall motions (traced by the extended diamond shape of the PV at low contour levels), 
a rotation signature is clearly seen in the form of two emission peaks at low 
velocities $\leq$ 1.5 km s$^{-1}$ (referred to as LV in the following): 
one blueshifted to the west, and one redshifted to the east. Figure 2c shows that 
this rotating LV C$^{17}$O emission
is (mainly) tracing the sides of the southern cavity carved by the outflow into the envelope. 
The sides are rotating 
in the same sense as the NH$_3$, C$^{18}$O, and HCO$^+$ cores
(Wiseman et al. 2001; Lee et al. 2006, 2014), i.e. with blueshifted gas in the west and redshifted gas in the east, 
as expected for envelope material swept-up and compressed by the outflow. 

Figure 2d shows that at higher velocities $\simeq$ 1.9--3.5 km s$^{-1}$ from 
systemic (denoted as HV in the following): (i) the E-W velocity gradient is still present, and (ii) the emission is 
definitely more compact ($\le 0\farcs5$).  
Similarly to other Class 0 disk studies (see e.g. Murillo et al. 2013, and references therein), the emission centroid positions in each 
channel were obtained from elliptical Gaussian fits 
in the $uv$ domain\footnote{We used the GILDAS $uv-fit$ task: 
the resulting error on centroid position is the function
of the channel signal-to-noise ratio and atmospheric seeing,
and is typically much smaller than the beam size.}.   
The results are plotted in
Fig.~\ref{fig:disk}. While
the centroids in the LV range fall on the southern cavity, as expected from
the channel maps, the centroids in the HV range 
move to the equatorial plane (within the uncertainties),  
which indicates an inner rotating equatorial disk of radius $\simeq 0\farcs2$ (90 AU). 
These findings are consistent with the outer radius of the flattened continuum disk imaged by Lee et al. (2014), 
who derived 120 AU using their ALMA map at 350 GHz. 

We compare in Fig.~\ref{fig:disk} the centroid measurements along the disk plane with Keplerian rotation curves 
and, as a reference, with a $\sim$ $r^{-1}$ rotation curve that conserves
specific angular momentum (pseudo-disk). 
Although a $\sim$ $r^{-1}$ trend is not ruled out, Fig.~\ref{fig:disk} shows that the present
measurements are consistent with keplerian rotation out to $\simeq 0\farcs2$ = 90 AU 
around a 0.3$\pm$0.1 $M_{\rm \odot}$ protostar.
This mass estimate agrees with the $0.2 M_{\odot}$ derived by Lee et al. (2006, 2014) 
from modelling the rotating infall kinematics of C$^{13}$O, C$^{18}$O, and HCO$^+$ on 
larger scales\footnote{The assumption of free-fall (used by Lee et al. 2014) overestimates 
the true infall speed in the presence of rotation (see e.g. Stahler et al. 1994, Sakai et al. 2014), 
and thus tends to underestimate the required central mass.} 
All these findings support the hypothesis of a rotating disk of radius 90 AU around the MM1 protostar, 
possibly in keplerian rotation inside this radius. 

\begin{figure}
\centerline{\includegraphics[angle=0,width=8.7cm]{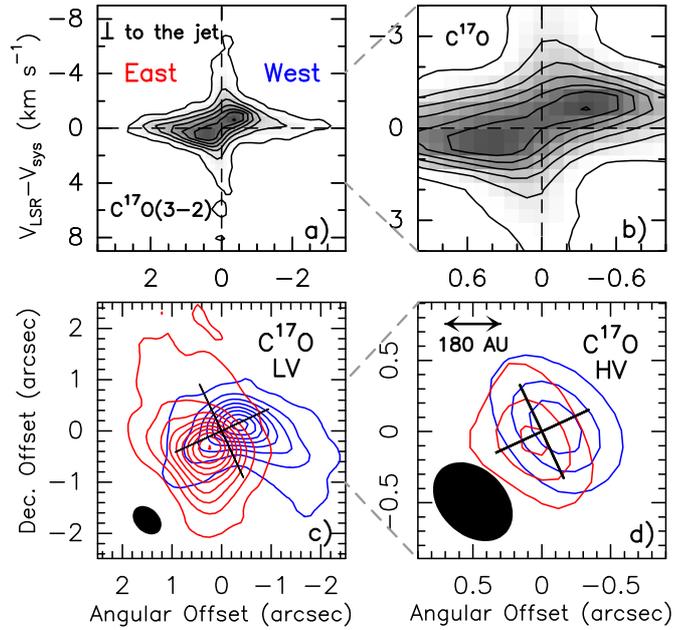}}
\caption{
{\it Panel a:} position-velocity (PV) cut of C$^{17}$O(3--2) 
perpendicular to the jet. First contour at 5$\sigma$ (0.30 K)
and steps of 20$\sigma$. Dashed lines mark $V_{\rm sys} = +1.3$ km s$^{-1}$ and the continuum peak MM1.
{\it Panel b:} Zoom-in of the C$^{17}$O(3--2) PV plot.
{\it Panel c:} Blue- and redshifted C$^{17}$O(3--2)
emission in the LV range ($\pm$0.6--1.5 km s$^{-1}$ from $V_{\rm sys}$), tracing the rotating outflow cavity.
The tilted black cross indicates the SiO jet direction (PA = 22$\degr$) and the equatorial plane.
First contour at 5$\sigma$ (15 mJy beam$^{-1}$ km s$^{-1}$), then steps of 7$\sigma$.
{\it Panel d:} same as panel (c) for the HV velocity range ($\pm$1.9--3.5 km s$^{-1}$ from systemic),
tracing the rotating inner disk. Note the smaller spatial scale in panels b and d.}
\label{pv}
\end{figure}

\begin{figure}
\centerline{\includegraphics[angle=0,width=7cm]{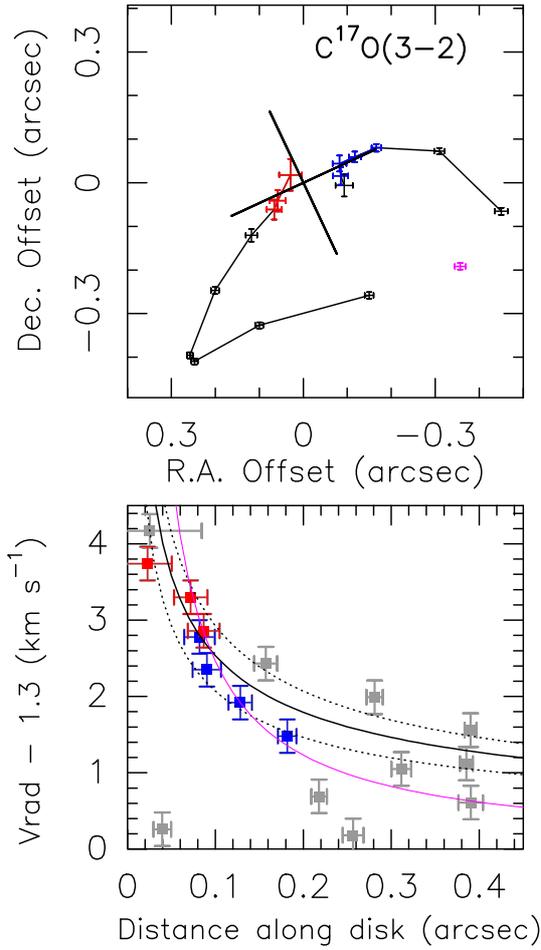}}
\caption{{\it Upper panel:} distribution of the C$^{17}$O(3--2) centroid positions 
(from fits in the $uv$ domain with 1$\sigma$ error bars) in the various velocity channels (Fig. A.1).
Red and blue datapoints denote the channels whose fitted centroids lie, within the error bars, on the equator.
{\it Bottom panel:} velocity shift from $V_{\rm sys}$ as a function of distance from the
protostar along the equator (PA = 122$\degr$).  
Red and blue datapoints denote the channels whose fitted centroids lie along the equator (see upper panel).
The assumed systemic velocity is +1.3 km s$^{-1}$ (see text). 
Grey points refer to LV emission tracing the C$^{17}$O envelope and cavity.
Black solid and dotted curves show keplerian rotation around a stellar mass of 0.3 $\pm 0.1 M_{\rm \odot}$.
The $r^{-1}$ curve for angular momentum conservation is plotted in magenta for comparison.}
\label{fig:disk}
\end{figure}

\subsection{C$^{34}$S: warped cavity and wide-angle flow}

\begin{figure}
\centerline{\includegraphics[angle=0,width=9cm]{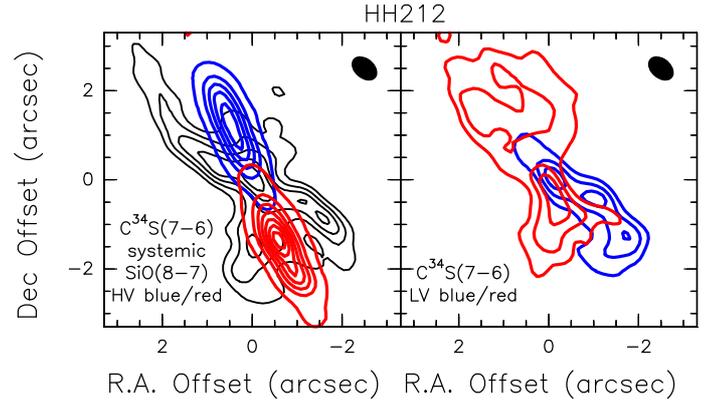}}
\caption{{\it Left:} SiO(8--7) channel maps 
at $V_{\rm sys} \pm$8 km s$^{-1}$ (blue and red) overlaid on top of C$^{34}$S(7--6) close to
systemic velocity (black). 
{\it Right:} C$^{34}$S(7--6) channel maps at low blue and redshifted velocities
$\sim$ $V_{\rm sys}$$\pm$0.9 km s$^{-1}$, showing a rotating wide-angle flow about the jet axis.  
SiO first contour at 5$\sigma$
(25 mJy beam$^{-1}$ km s$^{-1}$) and steps of 25$\sigma$.
C$^{34}$S first contours and steps correspond to 5$\sigma$
(15 mJy beam$^{-1}$ km s$^{-1}$).}
\label{maps2}
\end{figure}

While the bulk of the C$^{17}$O emission in Fig.1 (Middle) is tracing the protostellar envelope of FWHM
$\simeq$ 460 AU flattened in the equatorial plane,  C$^{34}$S is elongated along the outflow,
with little emission in the equatorial plane.
Given the high critical density of the C$^{34}$S(7--6) line ($\sim 9 \times 10^6$ cm$^{-3}$ between 10 K and 300 K, according
to the collisional rates of Lique et al. (2006)), 
and the envelope density inferred by Lee et al (2014), this suggests that CS is tracing a dense gas component 
more closely related to the primary jet or outflow.
Near systemic velocity, C$^{34}$S is imaging a biconical structure surrounding the SiO jet
and with MM1 at the vertex (see Fig.~\ref{maps2}). A surprising finding is the S-shaped warp seen in C$^{34}$S,
despite the very straight axial jet. Similar cavity asymmetries are predicted by MHD simulations during protostellar collapse
with a misaligned magnetic field and angular momentum vectors (e.g. Ciardi \& Hennebelle 2010);
hence the C$^{34}$S warp might be a remnant imprint of this initial configuration,
which has been invoked to explain the formation of large keplerian disks in Class 0 sources (Joos et al. 2012).
Sensitive polarisation measurements with ALMA will be crucial to test this hypothesis.

Figure~\ref{maps2} further indicates that the C$^{34}$S southern lobe is clearly rotating about the jet in 
the same sense as the C$^{17}$O cavity. 
However, Fig. A.2 shows that the structure C$^{34}$S is narrower than that of C$^{17}$O  
and becomes gradually more collimated towards the jet axis as the velocity increases, changing 
progressively from a biconical morphology near $V_{\rm sys}$ to a jet at $V_{\rm LSR}$--$V_{\rm sys}$ $\geq$ 2.5 
km s$^{-1}$.   
This suggests that the C$^{34}$S emission may be filling-in the swept-up extended cavity 
delineated by C$^{17}$O, and trace a rotating 
wide-angle flow with a nested onion-like velocity structure, highly reminiscent of that seen 
in the atomic jet from the T Tauri star DG Tau (Bacciotti et al. 2002). 
 
\section{Conclusions}

C$^{17}$O traces three different components that depends on the velocity
of its emission: the infalling envelope near systemic velocity, the rotating cavity in the LV range, 
and a rotating equatorial disk in the HV range with radius $\sim$ 0\farcs2 = 90 AU, which may be Keplerian
around a protostar of 0.3 $\pm 0.1 M_{\odot}$. 
On the other hand, the wide-angle CS flow suggests that disk winds may be present in this source.
The present results calls for more observations  
at high spatial and spectral resolutions of disk tracers to confirm the disk size and verify its keplerian nature.

\begin{acknowledgements}
We are grateful to D. Galli for helpful discussions and suggestions.
This paper makes use of the following ALMA data: ADS/JAO.ALMA\#2011.0.000647.S (PI: C. Codella). ALMA is a partnership of 
ESO (representing its member states), NSF (USA) and NINS (Japan), together with NRC (Canada) and NSC 
and ASIAA (Taiwan), in cooperation with the Republic of Chile. The Joint ALMA Observatory is operated by ESO, AUI/NRAO and NAOJ.
This work was partly supported by the PRIN INAF 2012 -- JEDI and by the 
Italian Ministero dell'Istruzione, Universit\`a e Ricerca through the grant Progetti Premiali 2012 -- iALMA.
LP has received funding from the European Union Seventh Framework Programme (FP7/2007-2013) under grant agreement n. 267251.
\end{acknowledgements}

\clearpage

\appendix

\section{C$^{17}$O, C$^{34}$S, and CH$_3$OH kinematics}

We report in Figs. A.1 and A.2 the channel maps of the 
C$^{17}$O(3--2) and C$^{34}$S(7--6) 
(continuum-subtracted) emissions towards HH212 that were analysed and discussed in
the main text.

\begin{figure*}
\centerline{\includegraphics[angle=0,width=16cm]{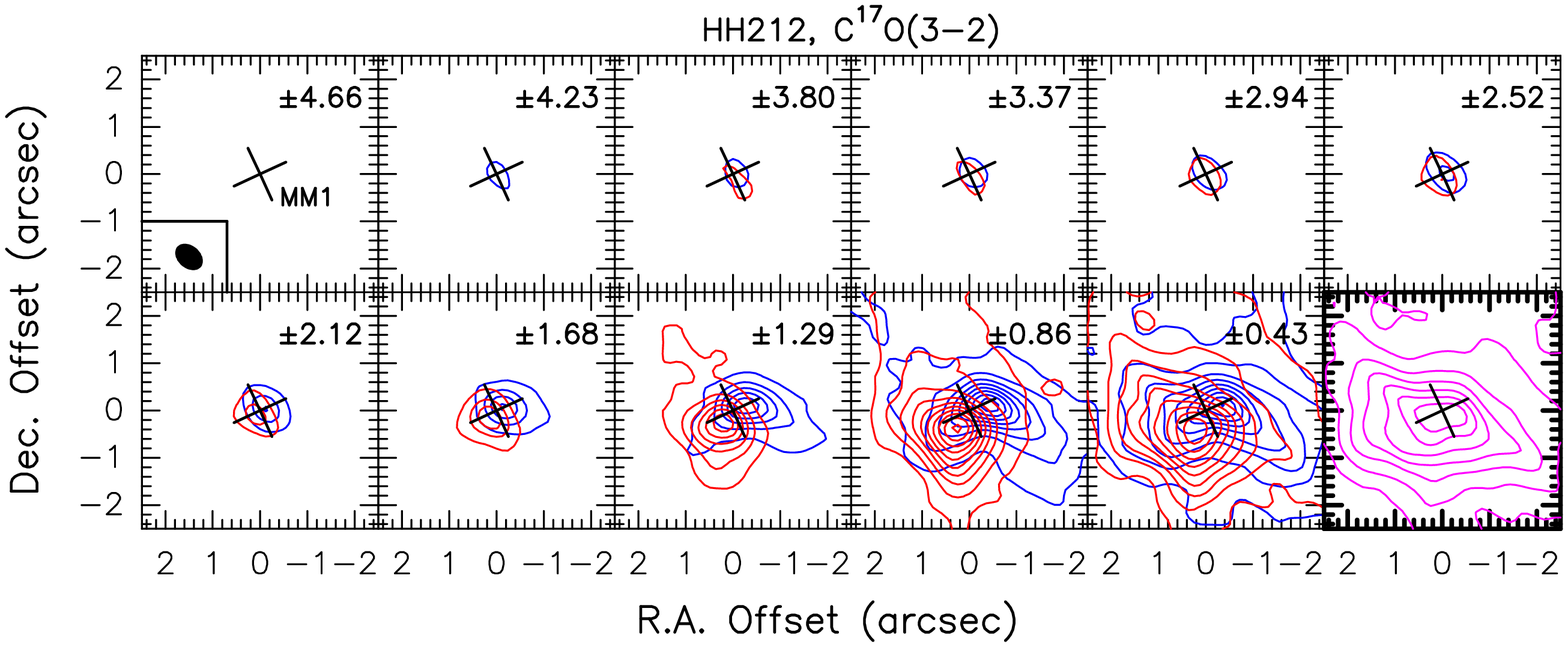}}
\caption{Channel maps of the C$^{17}$O(3--2) 
(continuum-subtracted) emissions towards HH212.
The bottom-right panel shows in magenta the 0.43 km s$^{-1}$
wide channel (centred on +1.13 km s$^{-1}$) where emission
is most spatially extended. The other panels superposed in blue and red contours are the channel
maps at symmetric blue/red velocity offsets from this central channel, with the velocity
shift given (in km s$^{-1}$) in the upper-right corner.
The black cross (inclined to show the SiO jet direction)
indicates the position of the MM1 continuum source.
The ellipse in the top-left panel shows
the ALMA synthesised beam (HPBW): $0\farcs65\times0\farcs47$ (PA = 49$\degr$).
First contours and steps correspond to 5$\sigma$ (15 mJy beam $^{-1}$ km s$^{-1}$).}
\label{fig:c17o-channels}
\end{figure*}

\begin{figure*}
\centerline{\includegraphics[angle=0,width=16cm]{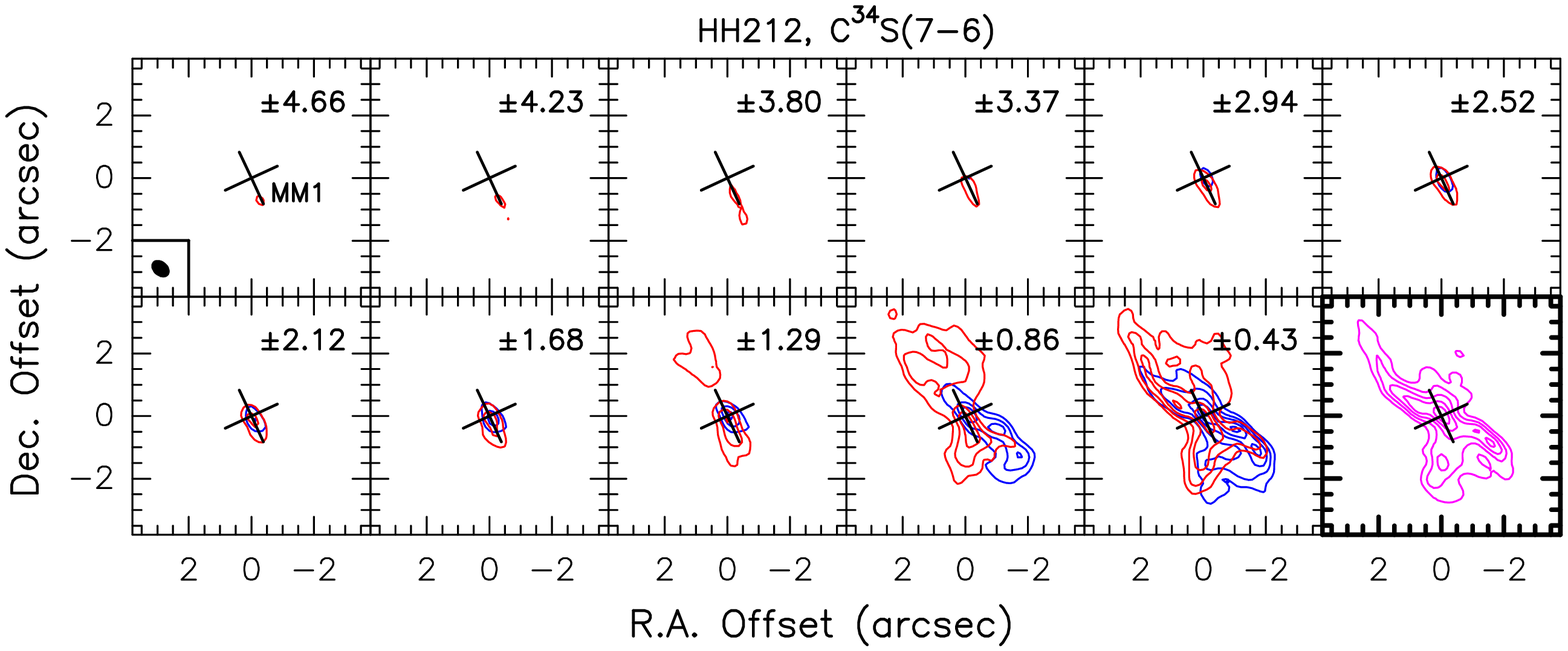}}
\caption{Channel maps of the C$^{34}$S(7--6) blue- and redshifted
(continuum-subtracted) emissions towards HH212.
The bottom-right panel shows in magenta the 0.43 km s$^{-1}$
wide channel (centred on +1.42 km s$^{-1}$) where emission
is the most spatially extended. The other panels superposed in blue and red contours are the channel
maps at symmetric blue and red velocity offsets from this central channel, with the velocity
shift given (in km s$^{-1}$) in the upper-right corner.
Symbols are the same as in Fig. A.1.
First contours and steps correspond to 5$\sigma$ (15 mJy beam $^{-1}$ km s$^{-1}$).}
\label{fig:c34s-channels}
\end{figure*}
 
\end{document}